\input{epsf}
\documentstyle[twocolumn,aps,prl,floats]{revtex} 
\tighten

\begin{document}
\def \lleq {\lower0.9ex\hbox{ $\buildrel < \over \sim$} ~}
\def \ggeq {\lower0.9ex\hbox{ $\buildrel > \over \sim$} ~}
\def\beq{\begin{equation}}
\def\eeq{\end{equation}}
\def\ber{\begin{eqnarray}}
\def\eer{\end{eqnarray}}
\def \apl {ApJ, }
\def \aps {ApJS, }
\def \pd {Phys. Rev. D, }
\def \prl {Phys. Rev. Lett., }
\def \pl {Phys. Lett., }
\def \np {Nucl. Phys., }
\def \l {\Lambda}

\draft
\twocolumn[\hsize\textwidth\columnwidth\hsize\csname @twocolumnfalse\endcsname
\title{A New Cosmological Model of Quintessence and Dark Matter}
\author {Varun Sahni$^{1,*}$ and Limin Wang $^{2,\dagger}$}
\address {$^1$Inter-University Centre for Astronomy \& Astrophysics,
Post Bag 4, Pune 411007, India} 
\address{$^2$Department of Physics, 538 West 120$^{\rm th}$ Street, Columbia University, New York NY 10027, USA}

\date{\today}

\maketitle

\begin{abstract}
We propose a new class of quintessence models in which
late times oscillations of
a scalar field give rise to an effective equation of state which 
can be negative
and hence drive the observed acceleration of the universe.
Our ansatz provides a unified picture of
quintessence and  a new form of dark matter we call {\em Frustrated Cold 
Dark Matter} (FCDM).  FCDM inhibits gravitational clustering on small 
scales and could provide a natural resolution to the core density problem
for disc galaxy halos.
Since the quintessence field rolls towards a small value,
constraints on slow-roll
quintessence models are safely circumvented in our model.
\end{abstract}

]

\narrowtext

The recent discovery that type Ia high redshift supernovae are fainter
than they would be in an Einstein-de Sitter universe suggests that
the universe may be accelerating, 
fuelled perhaps by a cosmological constant
or some other field possessing long range `repulsive' effects
\cite{perl,riess}.
The acceleration of the universe is related to the equation of state of 
matter through the Einstein equation
\beq
\frac{\ddot a}{a} = -\frac{4\pi G}{3} \big\lbrack\rho_c + 
\rho_X(1+3w_X)\big\rbrack
\label{eq:2}
\eeq
for cold matter $\rho_c$ and X-matter
with equation of state $P_X = w\rho_X$. Clearly a necessary (but not sufficient) condition for the universe to accelerate is $w_X < -1/3$. In other words
the equation of state of X-matter must violate the strong energy condition
(SEC) so that $\rho_X + 3P_X < 0$. 
Investigations
of cosmological models with $\Omega_m + \Omega_X \simeq 1$ have demonstrated
that these models outperform most others in predicting the correct form
for the large scale clustering spectrum, accounting for CMB anisotropies
on large and intermediate angular scales and providing excellent agreement
with the luminosity-distance relation obtained from observations of 
high redshift 
supernovae \cite{wang}. In addition, flat models are compelling
from a theoretical viewpoint since they agree with generic predictions made
by the inflationary scenario.

The literature describing phenomenological forms of matter violating the SEC
is vast (see \cite{ss99} for a recent review). Nevertheless two kinds of
matter have been singled out in recent times as being of special interest:

\begin{enumerate}
\item A cosmological constant $P_X = -\rho_X$ ($w_X = -1$), 
$\l \equiv \rho_X/8\pi G.$

\item  A scalar field rolling down a potential $V(\phi)$.
\end{enumerate}

For fields rolling sufficiently slowly
${\ddot \phi} \simeq 0$ and $T_{ik} \simeq V(\phi) g_{ik}$, so that $V(\phi)$
plays the role of a time-dependent $\l$-term. Although appealing, models
with the simplest potentials including 
$V \propto m^2\phi^2$ run into problems similar to 
those encountered by a cosmological constant. The enormous overdamping of the 
scalar field equation during radiation and matter dominated epochs causes 
$V(\phi)$ to remain unchanged virtually from the Planck
epoch $z_{pl} \sim 10^{19}$ to $z \sim 2$ \cite{frieman95} resulting in an
enormous difference in the scalar field energy density and that of 
matter/radiation at early times. This leads to a fine tuning problem:
the relative values of $\rho_\phi$ and $\rho_m$ must be set to very high
levels of accuracy $(\rho_\phi/\rho_m)_{\rm initial} \sim 10^{-123}$
in order to ensure $\rho_\phi/\rho_m \sim 1$ at 
precisely the present epoch. 

A more reasonable
assumption might be if the energy density in the $\phi$-field were 
comparable to that of radiation at very early times -- say at the end of
inflation \cite{tracker}. 
This might even be expected if the $\phi$-field were to be an
inflationary relic, its energy set by an equipartition ansatz. However
for the $\phi$-field to remain subdominant until recently its energy density must decrease rapidly at early times. Such behaviour clearly cannot arise
for polynomial potentials $V(\phi) \propto \phi^p, ~0 < p \lleq few$, 
for which $\rho_\phi$ will rapidly dominate the total density 
resulting in a colossal
$\l$-term today if $\rho_\phi \sim \rho_{\rm rad}$ initially. Fortunately there do exist families of potentials for which
the behaviour of $\rho_\phi$ is more flexible. 
To illustrate this, consider a minimally coupled scalar field
rolling down the potential 
\beq
V(\phi) = V_0(\cosh{\lambda\phi} - 1)^p.
\label{eq:pot1}
\eeq
$V(\phi)$ has asymptotic forms:
\ber
&V(\phi) ~&\simeq ~\tilde{V}_0e^{-p\lambda \phi} ~~{\rm for} ~
 \vert\lambda\phi\vert \gg 1
~(\phi < 0),
\label{eq:exp1}\\
&V(\phi) ~&\simeq ~\tilde{V}_0 (\lambda\phi)^{2p} ~~{\rm for}
~\vert\lambda\phi\vert \ll 1
\label{eq:exp2}
\eer
where $\tilde{V}_0 = V_0/2^p$. Scalar field models with the potential
$V(\phi) \propto e^{-p\lambda \phi}$ have the attractive property that
the energy density in $\phi$ tracks the the radiation/matter
component as long as the value of $\phi$ is large and negative, so that
\cite{exponential}:
\beq
\frac{\rho_\phi}{\rho_{B} + \rho_\phi} = \frac{3(1 + w_B)}{p^2\lambda^2}
\label{eq:ratra}
\eeq
($w_B = 0, ~1/3$ respectively for dust, radiation).
During later times the form of $V(\phi)$ changes to a power law (\ref{eq:exp2})
resulting in rapid oscillations of  $\phi$ about $\phi = 0$. The change
in the form of the scalar field potential is accompanied by an 
important change in
the equation of state of the scalar field. 
As long as $V(\phi)$ is described by (\ref{eq:exp1}), the kinetic energy of 
the scalar remains larger than its potential energy
 $\frac{1}{2}{\dot\phi}^2 > V(\phi)$ and
the scalar field equation of state mimicks background matter 
$w_\phi \simeq w_B$.
However during the oscillatory phase $\langle \frac{1}{2}{\dot\phi}^2\rangle$
can become smaller than $\langle V(\phi)\rangle$, the virial theorem then
gives the following expression for the mean equation of state \cite{turner}
\beq
\langle w_\phi\rangle = \bigg\langle \frac{\frac{1}{2}{\dot\phi^2} - V(\phi)}
{\frac{1}{2}{\dot\phi^2} + V(\phi)} \bigg\rangle = \frac{p - 1}{p + 1}.
\label{eq:w}
\eeq
The corresponding value of the scalar field density and expansion factor
is given by
\ber
&\rho_\phi \propto a^{-3(1+w_\phi)}
\label{eq:rho}\\
&a \propto t^c, ~ c = \frac{2}{3}(1+\langle w_\phi\rangle)^{-1}.
\label{eq:expansion}
\eer
 From (\ref{eq:w}), (\ref{eq:rho}) \& (\ref{eq:expansion}) 
we find that the mean equation of state, 
the scalar field density and the expansion rate of the universe
depend sensitively upon the value of the parameter $p$ in the potential 
(\ref{eq:pot1}).
Three values of $p$ should be singled out for particular attention since 
they give rise to cosmologically interesting solutions:

\begin{enumerate}
\item $p = 1$: In this case the scalar field equation of state behaves like
that of pressureless (cold) matter or dust $\langle w_\phi\rangle \simeq 0$,
a scalar field potential with this value of $p$ could therefore play the role of
cold dark matter (CDM) in the universe.

\item $p = 1/2$: This results in $\langle w_\phi\rangle
\simeq -1/3$ and $\rho_\phi \propto a^{-2}$. 
This choice of the parameter $p$ 
leads to a `coasting' form for the scale factor at late
times: $a(t) \propto t$. A flat universe under the influence of the potential
$V(\phi) = V_0(\cosh{\lambda\phi} - 1)^{1/2}$
will therefore have exactly the same expansion properties
as an open universe without being plagued by the `omega problem' !
(See \cite{open} for related scenarios.)

\item Smaller values $p < 1/2$ lead to $\langle w_\phi\rangle < -1/3$.
 From (\ref{eq:rho}) we find that
the scalar field density falls off 
slower than either radiation ($\rho_r \propto a^{-4}$) or
 cold matter ($\rho_m \propto a^{-3}$). 
The scalar field therefore
dominates the mass density in the universe at late times 
leading to accelerated expansion according to
$a \propto t^c, ~~ c = 2/3(1+\langle w_\phi\rangle) > 1$.
The epoch of scalar field dominance commences at the cosmological redshift
$1+z_{\star} = (\Omega_\phi/\Omega_m)^{-1/3\langle w_\phi\rangle}$.
\end{enumerate}

We therefore find that a scalar field with the potential (\ref{eq:pot1}) 
might serve as 
a good candidate for quintessence. Figure 1 confirms this by
showing the density parameter $\Omega_\phi$ as a function of
the cosmological scale factor. We find that the ratio 
$\Omega_\phi/\Omega_B$ remains 
approximately constant
during the prolonged epoch of radiation/matter domination as 
the $\phi$-field tracks the dominant radiation/matter component.
($\Omega_\phi/\Omega_B < 0.2$ is necessary in order to satisfy nucleosynthesis
constraints.) At the end of the matter dominated epoch
$\Omega_\phi/\Omega_B$ begins to grow as the scalar field equation of state
turns negative in response to rapid oscillations of $\phi$. 
As in earlier tracker quintessence models \cite{tracker} present-day 
values $\Omega_\phi
\simeq 0.7$, $\Omega_m \simeq 0.3$, consistent with 
supernovae \cite{perl,riess} and other observations \cite{wang}, 
can be obtained for a large class of
initial conditions. 
(General potentials leading to oscillatory quintessence must satisfy
$\langle V - \phi V'\rangle > 0$
where angular brackets denote the time average over a single oscillation
\cite{damour}.
We assume that the quintessence field couples very weakly to other matter
fields so that rapid oscillations of $\phi$ do not result in particle
production of the kind associated with `preheating'.)
Our potential belongs to the general category of exponential potentials
which are frequently encountered in field theory \cite{field},
condensed matter physics \cite{pot} and as solutions
to the non-perturbative renormalization group equations \cite{branchina}.
\begin{figure}
\epsfxsize=3.3 in \epsfbox{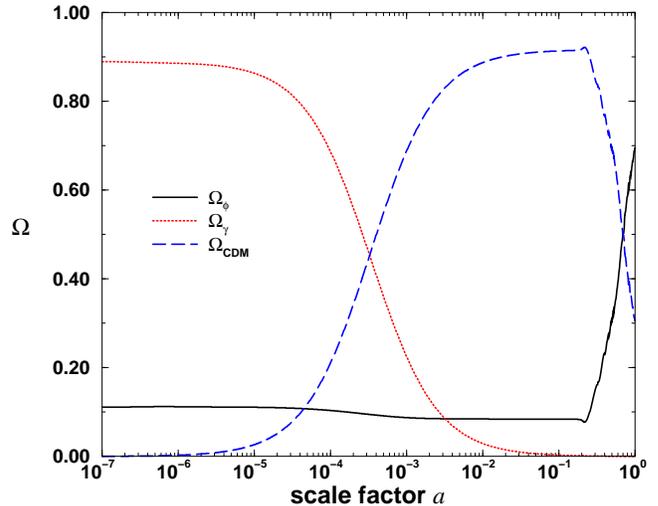}
\caption{
The evolution of 
the dimensionless density parameter for the quintessence
field $\Omega_\phi$ (solid line) 
is shown for the potential (\ref{eq:pot1}) with $p = 0.2$.
Matter (dashed line) and radiation densities (dotted line)
are also shown. For $z \gg 1$ the quintessence
field satisfies the tracker solution (\ref{eq:ratra}) 
and contributes a fixed fraction to
the total density of background matter/radiation. 
At later times $z \lleq 2$ scalar field oscillations 
commence and the density in the quintessence component rapidly dominates
the mass density of the universe leading to $\Omega_\phi \sim 0.7$ today.
In this figure the parameters of the quintessence model used are:
${\tilde V}_0 = 8\times 10^{-9} m_{Pl}^2 Mpc^{-2}$, 
$\lambda = 30 m_{Pl}^{-1}$. }
\end{figure}

In figure 2 we compare the
redshift dependence of the luminosity distance for a specific realisation
of our quintessence model with $d_L$ 
obtained from supernovae observations.

\vspace{.4cm}
\epsfxsize=3.3 in \epsfbox{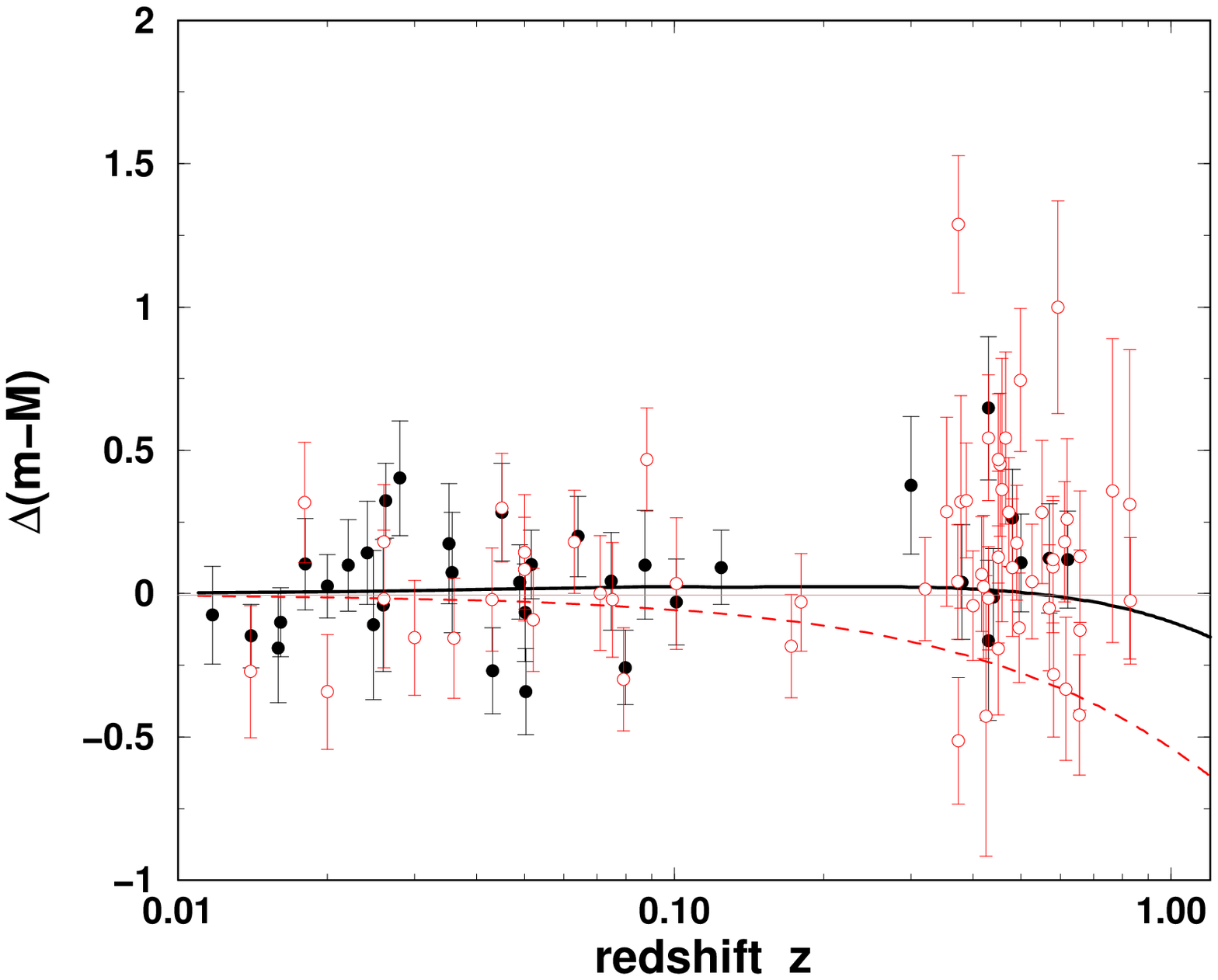}
\vspace{0cm}
{FIG. 2. {\small{The luminosity distance {\it vs.} redshift for the model
shown in fig. 1 (solid line).  The dashed line is the standard CDM model
(shown for comparison) and the horizontal line corresponds to the fiducial
empty Milne universe $\Omega_m \rightarrow 0$.  
The filled circles show supernovae data from High-Z
Supernova Search Team\cite{riess} and the opaque circles show data from 
Supernova Cosmology Project\cite{perl}. 
The low-z supernovae are from the Calan-Tololo sample.
}}} \\

We would also like to draw attention to the possibility of a unified picture
of quintessence and cold dark matter in which both components are described by
a pair of scalar fields evolving under the action of 
the potential (\ref{eq:pot1}) but with different values of the exponent $p$:
\beq
V(\phi,\psi) = V_{\phi}(\cosh{\lambda_{\phi}\phi} - 1)^{p_{\phi}}
   + V_{\psi}(\cosh{\lambda_{\psi}\psi} - 1)^{p_{\psi}}
\label{eq:potCDM}
\eeq
$p_{\psi} = 1$ in the case of CDM
and $p_{\phi} \lleq 0.5$ in the case of quintessence. This approach 
(along to the lines suggested by \cite{zlatev99})
ameliorates the `coincidence problem' between
dark matter (CDM) and quintessence which arises in standard
cosmology. It also significantly reduces the discrepancy between the
present value of $\rho_m/\rho_r$ and that at the end of Inflation.
Figure 3 shows a working example.  

It is interesting that the CDM particle in this scenario can be ultra-light,
its mass ($m^2 = V'' \sim H^2(t_*)$) is related to the epoch $t_*$ when 
$\psi$ begins to oscillate and its Compton wavelength $\lambda_c = m^{-1}$
can easily be of order a kilo-parsec or smaller.
(In the cosmological model illustrated
in figure 3 the CDM field $\psi$ begins
to oscillate at $z\approx 10^{5}$ so that $m^{-1} \simeq 228$ parsec.)
Cold dark matter made up of a condensate of ultra-light particles 
would be frustrated in its attempts to
cluster on scales smaller than $\lambda_c$ because of the uncertainty
principle, the resulting {\em Frustrated Cold Dark Matter} model (FCDM)
might provide a natural explanation for two major difficulties
faced by the standard CDM scenario. 
(i) The dearth of halo dwarf galaxies: the number of dwarf's in the
local group is an order of magnitude smaller than predicted by N-body 
simulations of SCDM \cite{moore99a}. 
(ii) The discrepency between observed shapes of galaxy 
rotation curves and simulated dark matter halos.
Recent observations of low surface brightness (LSB) galaxies
show them to possess rotation curves which indicate a constant mass 
density in the central core region. These observations are difficult to
accomodate within the SCDM model since high resolution N-body simulations
of SCDM halos indicate a cuspy central density profile having the form
$\rho \propto r^{-1.5}$ in the core region
 \cite{moore99b,navarro99,ghigna99,SpergelSteinh99,KamionLiddle99}.

\vspace{.4cm}
\epsfxsize=3.3 in \epsfbox{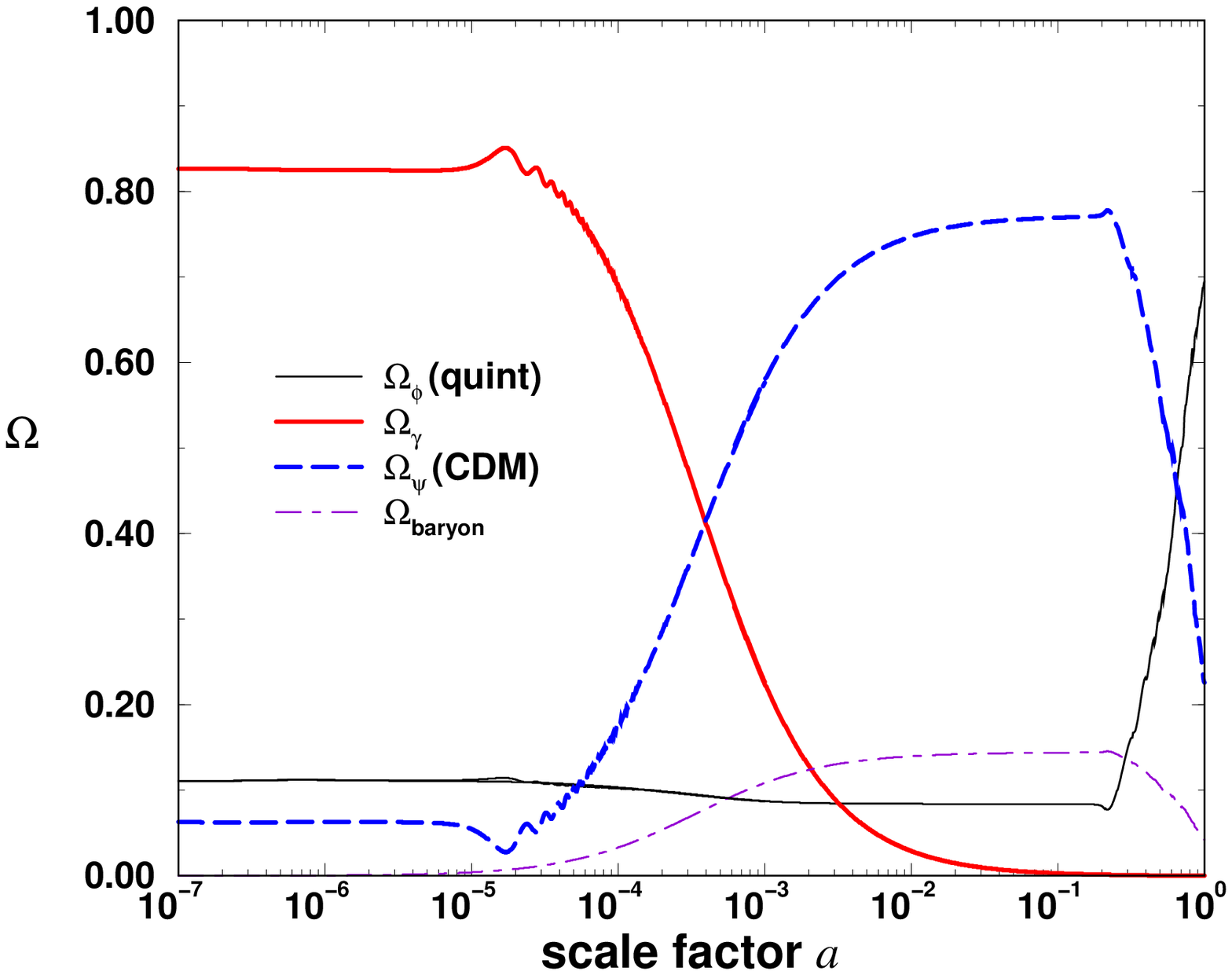}
\vspace{0cm}
{FIG. 3. {\small{The evolution of the dimensionless density parameter for
the CDM field $\Omega_{\psi}$ (dashed line) and quintessence field
$\Omega_{\phi}$ (thin solid line).  Baryon (dash-dotted line) and radiation
densities (thick solid line) are also shown.  The parameters for $\phi$
have been given in the caption of figure 1.  The parameters for $\psi$
are:
$p_{\psi}=1$,
${\tilde V}_{\psi} = 3\times 10^{5} m_{Pl}^2 Mpc^{-2}$,
$\lambda_{\psi} = 8 m_{Pl}^{-1}$.
}}} \\

It is both interesting and revealing 
that a physical mechanism suppressing small scale
clustering arises in the FCDM model at the purely classical level.
To demonstrate this we note that once $\psi$ begins oscillating 
$V(\psi)$ acquires the form $V(\psi) \simeq V_0\lbrack\frac{1}{2}\lambda^2\psi^2
+ \frac{1}{24}\lambda^4\psi^4\rbrack$. Using the relationship \cite{pv99}
$\langle{\dot\psi}^2\rangle = V_0\lbrack\lambda^2\langle\psi^2\rangle + 
\frac{\lambda^4}{6}\langle\psi^4\rangle\rbrack$ and assuming that the 
gradient energy is subdominant we obtain the following
expressions for the mean
pressure $p_\psi$ and the mean density $\rho_\psi$:
$p_\psi = \frac{V_0}{24}\lambda^4\langle\psi^4\rangle$, $\rho_\psi 
\simeq V_0\lambda^2\langle\psi^2\rangle$. The motion of $\psi$ is driven
mainly by the quadratic term $\lambda^2\psi^2$ (which also provides the
dominant contribution to the energy density). As a result
${\dot\psi} = \sqrt{V_0\lambda^2(\psi_0^2-\psi^2)}$, which can be
 used to establish
$\langle\psi^4\rangle = \frac{3}{2}\langle\psi^2\rangle^2$. 
Substituting the resulting expression for the speed of sound
$v_s^2 = dp_\psi/d\rho_\psi = \rho_\psi/8V_0$
into the Jeans length $\lambda_J \simeq \sqrt{v_s^2/2\pi G\rho_\psi}$, we get
\beq
\lambda_J \simeq \sqrt{\frac{m_{Pl}^2}{2V_0}} \equiv
\frac{\lambda}{\sqrt{2}}(\frac{m_{Pl}}{m}).
\eeq
We therefore find that the Jeans length is {\em larger} than the Compton 
wavelength 
($\lambda_J > m^{-1}$) if $\lambda > m_{Pl}^{-1}$ \cite{discussion}.
For the FCDM model illustrated in figure 3, one finds
 $\lambda_J \simeq$ 1 kpc. 
A lighter particle will possess a larger
Jeans length while $\lambda_J$ is smaller for a $\psi$-field which began
oscillating much before $t_{eq}$, a very massive field will resemble standard 
CDM.
By inhibiting gravitational clustering on scales  smaller than 
$\lambda_J \sim $ kpc,
FCDM is expected to
give rise to galaxy halos which are less centrally concentrated 
leading to better agreement between theory and observations.

Finally we note that quintessence-type potentials could also arise in 
particle physics models which invoke the Peccei-Quinn mechanism to
solve the strong CP problem in QCD. Consider for instance the following
simple modification to the symmetry breaking potential responsible for the 
axion
\beq
V(\phi) = \lambda(|\phi|^2 - \frac{f^2}{2})^2 + 
m^2f^2(1 - \cos{\theta})^p.
\label{eq:axion}
\eeq
The second term in (\ref{eq:axion}) when expanded about $\theta \equiv
{\rm arg}\langle\phi\rangle = 0$ acquires the form
$m^2f^2(\theta^2/2)^{p}$. Accordingly
rapid oscillations of the $\theta$ field about $\theta = 0$ now
give rise to an equation of state described by (\ref{eq:w}), resulting in
$\langle w_\theta\rangle \leq -1/3$ for $p \leq 1/2$.
(In the standard scenario
$p = 1$, $\langle w_\theta\rangle = 0$,
$m \rightarrow m_a$ is the axion mass and $f \rightarrow f_{PQ}$
is the Peccei-Quinn symmetry breaking scale.)
As a result an
axion-like scalar with $p \leq 1/2$ will
be a candidate for quintessence since its energy density will
diminish more slowly than that of either matter or radiation, 
leading to the dominance of the $\theta$-field at late times and the 
accelerated expansion of the universe.

It should be noted that motion under the action of potentials (\ref{eq:pot1}) 
\& (\ref{eq:axion})
is well defined even though for $p < 1/2$, $V'(\phi)$ is weakly singular
at $\phi = 0$. One can make $V(\phi)$ mathematically
more appealing by the field redefinition $\phi \to (\phi^2 + \phi_c^2)^{1/2}$,
$\phi_c \to 0$, this will not affect our results in any significant way.

It is also worth mentioning that $V'' < 0$ during the oscillatory stage
if $p < 1/2$. This is likely to affect very long wavelength fluctuations
in $\phi$ for the potential (\ref{eq:pot1}) since the scalar field begins
oscillating fairly recently in this case. 

Finally we would like to point out an important distinction which 
exists between the
quintessence models suggested by us and those of \cite{tracker}. 
In our models the quintessence field $\phi$ or $\theta$
oscillates about a {\em small} value at late times
(formally $\phi,\theta \rightarrow 0$ as $t \rightarrow \infty$). 
Additionally the potential $V(\phi)$ is not constraind to be flat
since the field does not have to slow-roll in order to give rise to a 
negative equation of state.
As a result quantum corrections which might be significant in quintessence 
models
in which the scalar field rolls down a flat potential 
to large values $\phi \ggeq m_{Pl}$
\cite{lyth} can safely 
be ignored in models of the kind discussed in the present paper.

We acknowledge the hospitality of the organisers of
the Santa Fe Summer Workshop on Structure
Formation and Dark Matter, June 28 - July 16, 1999 where this work was
initiated. 
We thank Lloyd Knox for pointing out that the Compton wavelength
of our CDM particle has an astrophysically interesting length scale
which subsequently led to many interesting discussions.
Useful discussions with Carlos Frenk, Marc Kamionkowski,
Somak Raychaudhury, Jim Peebles,
Rachel Somerville, Alexei Starobinsky and Neil Turok are also acknowledged. 
LW was supported by DoE grant DE-FG02-92ER40699.


\begin{references}
\bibitem{discussion}
We benefitted from discussions with Jim Peebles and Alexei
Starobinsky on this issue.
\end{references}
\end{document}